# In defense of real quantum theory

J. Finkelstein[*]

## Abstract

Two recent papers (Renou et al., arXiv:2101.10873, and Chen et al., arXiv:2103.08123) have indicated that complex numbers are necessary for quantum theory. This short note is a comment on their result.

Textbook quantum theory is formulated with complex numbers; for example, quantum states are taken to be represented by operators on a complex Hilbert space. Is that merely for mathematical convenience (as is the use of complex numbers in classical electromagnetism)? Or is there some aspect of quantum phenomena which *requires* the use of complex numbers?

Two recent papers have apparently given an affirmative answer to this second question. Ref. 1, entitled "Quantum physics needs complex numbers", has proposed an experiment for which a quantum theory based on real numbers (which I will call a "real quantum theory") would necessarily give different results than would standard "complex quantum theory". Ref. 2 ("Ruling out real-number description of quantum mechanics") reports results of this proposed experiment which agree with complex quantum theory, but disagree with real quantum theory.

On the other hand, both ref. 1 and ref. 2 also remind the reader of the construction of a real quantum theory which is guaranteed to give results identical with those of complex quantum theory: For any system, let $\rho$ represent the quantum state of that system in the usual complex theory; let $|\pm i\rangle = \{|1\rangle \pm i|0\rangle\}/2^{1/2}$ be two states of an auxiliary qubit, and define

$$\rho' = \tfrac{1}{2}\{\rho \otimes |+i\rangle(+i| + \rho^* \otimes |-i\rangle(-i|\} \qquad (1)$$

$\rho'$ is a real operator, and so can be considered as an operator on a real Hilbert space. Then (with similar definitions for other operators) this real quantum theory can be seen to be operationally equivalent to complex quantum theory.

---

[*]Lawrence Berkeley National Laboratory
 jlfinkelstein@sonic.net

At first sight, the preceding two paragraphs may seem to be in conflict: if eq. 1 leads to a real quantum theory equivalent to complex quantum theory, then how could there be an experiment which could distinguish between them? This apparent conflict is resolved when one realizes that ref. 1 only considered real quantum theories which subscribe to the "standard quantum formalism" (as they call it), in particular which respect the tensor-product rule (the rule which says that, for example, the Hilbert space for a system of two particles is the tensor product of the two Hilbert spaces of the individual particles). The real quantum theory which begins with eq. 1 is not among those which are ruled out by refs. 1 and 2, since it does not respect the tensor-product rule.

To avoid misunderstanding, let me state that I am not disputing anything in refs. 1 and 2. In fact, everything I have written above is a re-statement of points made in those two papers. In this note, I just want to raise the question of whether, when considering if a theory for quantum phenomena must involve complex numbers, one should disregard real quantum theories which do not respect the standard formalism of complex quantum theory.

If someone were to announce that she had a theory which, for example, rejected the tensor-product rule, one might worry, at least before her theory was specified more fully, that, for example, superluminal signaling might be allowed. For the real quantum theory which begins with eq. 1, there is no such worry. We know that this theory does not allow superluminal signaling because it is equivalent to complex quantum theory which does not allow superluminal signaling. More generally, although this real quantum theory does not respect the standard quantum formalism, since it is guaranteed to be observationally equivalent to complex quantum theory, it is not easy to see how anything could possibly go wrong.

In summary: refs. 1 and 2 do not rule out *any* real quantum theory (nor do they claim to); instead, they rule out real quantum theories which respect the formalism of the standard complex theory.